\documentclass[prb,twocolumn,aps,superscriptaddress,longbibliography]{revtex4-2}

\usepackage{bm}%
\usepackage{float}
\expandafter\ifx\csname package@font\endcsname\relax\else
 \expandafter\expandafter
 \expandafter\usepackage

There is also the cancel package:

 \expandafter\expandafter
 \expandafter{\csname package@font\endcsname}%
\fi

\usepackage{array}
\usepackage{amsmath,amssymb}
\usepackage{MnSymbol}
\usepackage{tikz-feynman,contour} 
\usetikzlibrary{shapes,arrows,positioning,automata,backgrounds,calc,er,patterns}
\usepackage{feynmf}
\tikzfeynmanset{compat=1.0.0} 

\usepackage{graphicx}
\usepackage{mathrsfs}
\usepackage{bm}
\usepackage{mathtools}
\usepackage{epstopdf}
\usepackage{hyperref}
\usepackage[normalem]{ulem}

\hypersetup{%
   pdfpagemode=None, %FullScreen,
   pdfstartpage=1,
   pdfmenubar=true,
   pdftoolbar=true,
   colorlinks = true,
   linkcolor=blue,
   citecolor=blue,
   urlcolor=blue,
   bookmarksopen=false
 }
 \usepackage{amsmath}
\usepackage{amsfonts}
\usepackage{mathtools}
\usepackage{graphicx}
\usepackage{fixme}
\usepackage{color}
\usepackage{cancel}

\def\bk{{\bf k}}
\def\bp{{\bf p}}
\def\bq{{\bf q}}
\def\bk{{\bf k}}

\def\bG{{\bf G}}

\begin{document}

%\begin{centering}
%\begin{Large}
%\bigskip \bigskip
%\end{Large}
%\title{Microscopic theory of an exciton in a Wigner crystal}
\title{Exciton interacting with the phonons of an electronic Wigner crystal }

%{Authors $^{1,2\ast}$}
%\bigskip
%\textit{
%$^{1}$Department of Physics and Astronomy, Aarhus University, Ny Munkegade, 8000 Aarhus C, Denmark. \\ 
%$^{2}$Shenzhen Institute for Quantum Science and Engineering and Department of Physics, Southern University of Science and Technology, Shenzhen 518055, China. \\ 
%xx}
%\bigskip

%$^\ast$To whom correspondence should be addressed; E-mail: bruungmb@phys.au.dk

%\end{centering}

%\bigskip
%\bigskip
\author{Jens Havgaard Nyhegn}
\affiliation{Center for Complex Quantum Systems, Department of Physics and Astronomy, Aarhus University, Ny Munkegade 120, DK-8000 Aarhus C, Denmark.}
\affiliation{Niels Bohr Institute, University of Copenhagen, 2100 Copenhagen, Denmark}
\affiliation{Kavli Institute for Theoretical Physics, University of California, Santa Barbara, California 93106-4030, USA}
\author{Esben Rohan Christensen}
\affiliation{Center for Complex Quantum Systems, Department of Physics and Astronomy, Aarhus University, Ny Munkegade 120, DK-8000 Aarhus C, Denmark.}
\affiliation{Accademia di Danimarca – The Danish Academy for Science and Art in Rome, \\
Via Omero 18, 00197 Roma RM, Italy.}
\author{Georg M. Bruun} 
\affiliation{Center for Complex Quantum Systems, Department of Physics and Astronomy, Aarhus University, Ny Munkegade 120, DK-8000 Aarhus C, Denmark.}
%xx}}

\begin{abstract}
With the advent of  atomically thin and tunable van der Waals materials, a two-dimensional  electronic Wigner crystal has recently been observed. The smoking gun signal was the appearance of an umklapp branch in optical exciton spectroscopy
coming from the  periodic potential generated by the Wigner 
crystal assumed to be static. Vibrations of the Wigner crystal, however, lead to a gapless phonon spectrum, which may affect the 
exciton spectrum. To explore this, we develop a  field-theoretical description of an 
exciton interacting with electrons forming  a Wigner crystal including  
the coupling to the phonons. We show that the
importance of the exciton-phonon coupling 
scales with the exciton-electron interaction strength relative  to the typical phonon energy squared.
The motion of the exciton leads to two kinds of scattering processes,
where the exciton emits a phonon either staying within the same Bloch band (intraband scattering) or changing its band (interband scattering). Using a 
nonperturbative self-consistent Born approximation, we demonstrate that these scattering processes  
lead to the formation of quasiparticles (polarons) consisting of the exciton in Bloch states dressed by Wigner crystal phonons. 
The energy shift and damping of these  polarons depend on the electron density in a nontrivial way since it affects both the 
exciton-phonon interaction strength as well as  the phonon and exciton spectra. In particular, the damping is strongly affected by whether the 
polaron energy is inside the gapless phonon scattering continuum or not. Using these results, 
we finally analyze their effects on the observed spectral properties of the 
exciton. 
 \end{abstract}

%Recent experimental advancements using layers of Transition metal dichalcogenides (TMDs) have reignited interest in this exotic state \cite{}. Typically a strong $B$-field is needed to enter this phase because it quenches the kinetic energy allowing the dynamics to be dominated by the Coulomb interactions. To get around this
 
\date{\today}
\maketitle
The competition between the kinetic energy of electrons and their Coulomb repulsion determines the electronic properties of many materials. 
In most cases, the kinetic energy dominates so that the electrons can 
be accurately described by Fermi liquid theory but as first predicted by Wigner~\cite{Wigner1934}, 
 the electrons may instead form a crystal structure when the Coulomb interaction dominates.
 For a two-dimensional (2D) electron gas, quantum Monte Carlo calculations show that a Wigner crystal is formed when the 
 ratio of the typical Coulomb interaction energy 
 over the kinetic energy $r_s=m^*_e e^2/(4\pi \epsilon_0 \epsilon_r \hbar^2 \sqrt{\pi n_e})$ exceeds approximately $30$~\cite{Drummond2009}.
Here $n_e$ is the density of the electrons, $m_e^*$ is their effective mass, and  $\epsilon_r$ is the dielectric constant. With the 
advent of transition-metal dichalcogenides (TMDs), which are atomically thin materials with a relatively large $m_e^*$, small $\epsilon_r$ due to the 
reduced screening, and a 
tunable electron density $n_e$~\cite{Wang2018}, this condition was finally overcome~\cite{Zhou2021,Smolenski2021}. In these experiments, 
the formation of a Wigner crystal was identified by the appearance of an umklapp  branch in the exciton spectrum as it 
is subject to a 
periodic potential created by the electrons. Optical exciton spectroscopy and umklapp scattering were subsequently used to 
explore the melting of the Wigner crystal for increasing electron density~\cite{Sung2025}, which  was later imaged 
directly using scanning tunneling microscopy~\cite{Xiang2025}. Finally, the presence  of a pinning mode in terahertz spectroscopy has recently been argued to 
arise from a Wigner crystal~\cite{chen2025terahertzelectrodynamicszerofieldwigner}.

So far, the theoretical analysis of an exciton in an electronic Wigner crystal has been based on 
mean-field theory where the dynamics of the electrons is ignored. In this approach,  the Wigner crystal  produces a static periodic potential felt by the excitons giving rise to a Bloch band structure~\cite{Yuya2021,Smolenski2021}. The small group velocity of this  band structure   at the Brillouin zone (BZ) edges was recently predicted to give rise to a reduction in the exciton 
diffusion coefficient~\cite{erkensten2025impactelectronwignercrystal}. Recently, a purely electronic theory for exciton-polaron formation was  presented~\cite{pichler2025purelyelectronicmodelexcitonpolaron}. 
The motion of the 
exciton, however, will excite vibrations (phonons) in the Wigner crystal, and the effects of this have so far 
been ignored. This must in general be expected to change the exciton spectrum in particular because  the phonon spectrum of a Wigner crystal is 
gapless, in analogy  
with the formation of attractive and repulsive polarons when the exciton is coupled to particle-hole excitations in an electron gas~\cite{massignan2026}.  

In this paper, we address this question and analyze the properties of an exciton interacting with electrons in  a
Wigner crystal taking into account the formation of phonons. 
Using a field theoretical description, we show that the exciton can emit and absorb phonons in two different ways where it either 
stays  within the same Bloch band (intraband scattering) or where it changes its band (interband scattering). We give a 
microscopic expression for the strength of the exciton-phonon interaction and develop a 
 self-consistent Born approximation to describe these scattering processes nonperturbatively. The exciton-phonon scattering leads to the 
formation of Bloch band polarons consisting of the exciton dressed by Wigner crystal phonons in 
close analogy with the original polaron model~\cite{Pekar}. We show that the energy and damping of these polarons depend critically 
on the electron density, since it affects the exciton-phonon interaction strength and also the spectra of the phonons 
and the excitons. This in particular 
means that the polaron energy is inside the gapless phonon scattering continuum only for certain intermediate electron densities leading to 
strong damping. We finally analyze how these results affect the observed optical exciton spectrum.

%What methodologies are best suited to treat this problem for different couplings? \esben{To be improved...}\\

\begin{figure}[H]
\centering
  \includegraphics[width=0.8\columnwidth]{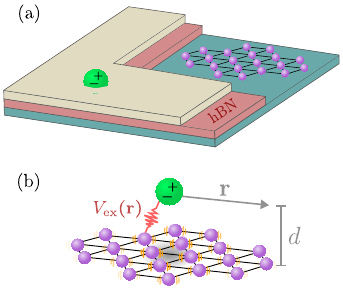}
\caption{(a) An exciton in the top layer interacts with the electrons forming a Wigner crystal in the lower layer. (b) This  leads to the formation of lattice vibrations (phonons).}
\label{fig.drawing}
\end{figure}

\section{Introduction} %{\it System.---}%We consider our system as a big 2D box of volume $\mathcal{V}$.
Inspired by the experimental setups with TMD materials, we explore the system illustrated in Fig. \ref{fig.drawing}(a). It consists of a top 2D layer containing an exciton and 
a bottom layer with electrons forming a triangular Wigner crystal (WC), separated by a middle insulating hexagonal boron
nitride (hBN) layer of thickness $d$ prohibiting electrons from tunneling between the bottom and the top layer.
Considering such a bilayer setup  avoids any complications arising from the exciton consisting of the same electrons as those forming the WC, allowing us to focus on the effects of the phonons. It also 
makes it possible to accurately describe  the exciton-electron interaction as that between a charge and a dipole; see below.  
% between the WC and the top layer. 
The equilibrium lattice positions ${\mathbf R}_i^0$ of the electrons forming the WC are 
spanned by the vectors $\mathbf{a}_1=a(\sqrt{3},1)/2$ and $\mathbf{a}_2=a(-\sqrt{3},1)/2$
where $a$ is the lattice constant. This gives the reciprocal lattice (RL) vectors ${\mathbf G}$ spanned by 
  $\mathbf{b}_1=(1/\sqrt{3},1)2\pi/a$ and $\mathbf{b}_2=(-1/\sqrt{3},1)2\pi/a$.

%With the exciton being charge neutral, the 
%long range  interaction between it and the WC is the dipole-charge coupling \cite{sidler2017}. Illustrated in Fig. \ref{fig.drawing}(b), this produces an attractive force between the electrons in the WC and the exciton creating umklapp scattering due to the periodic potential along with lattice deformations, phonons.  \\ \indent

%\subsection{Solving the Bloch problem}
%We want to study the physics of an exciton coupled to the lattice of electrons. We imagine that the exciton exists in its own layer separated from the layer of the Wigner crystal by a distance $d$, so that it has coordinates $\mathbf{r}_X=(x_X,y_X,d)^T$. Then the exciton-electron interaction potential must have the form: 
The exciton in the top layer interacts with  the electrons in the WC giving rise to umklapp scattering as well as phonon emission/absorption as illustrated in Fig.~\ref{fig.drawing}(b),
which is the main focus of the present paper.  
While calculating the exciton-electron interaction for short distances in general requires solving the three-body electron-electron-hole 
problem~\cite{Emfkin2021,fey2020}, the interaction  approaches the attractive charge-dipole potential
\begin{equation}
    V_\text{ex}(\mathbf{r}) = -\frac{\kappa}{(d^2 + r^2)^2}, %= -\frac{\alpha e^2}{(\epsilon_r \epsilon_0)^2\left(d^2 + r_{||}^2 \right)^2},
     \label{potentialfull}
\end{equation}
 for distances larger than the exciton radius $a_x$. Here,  $\kappa=\alpha e^{2}/(\epsilon_{r}\epsilon_{0})^{2}$ with $\alpha$  the polarizability of the exciton and $\epsilon_0\epsilon_r$ 
  the vacuum permittivity scaled by the material-dependent factor $\epsilon_r$, and 
 $\mathbf{r}=(x,y)$ is the 2D separation between the electron and the exciton in the  plane defined by the layers.

Following the standard procedure for deriving the Fr\"ohlich electron-phonon Hamiltonian for crystals~\cite{Bruus04}, 
we show in Appendix~\ref{App.Ham} that the Hamiltonian describing the system can be written as %$\hat H=\hat H_0+\hat H_\text{int}$ with
\begin{gather}
\hat H= \sum_{\mathbf{k}} \epsilon_{\mathbf k} \hat x^\dagger_\mathbf{k}\hat x_\mathbf{k}+\sum_{\substack{\mathbf{q}\in \text{BZ}\\ \lambda}}  \omega_{\mathbf{q}\lambda} \hat b^\dagger_{\mathbf{q}\lambda} \hat b_{\mathbf{q}\lambda}+\sum_{\substack{\mathbf k\\{\mathbf G}\in \text{RL}}}V_\text{B}({\mathbf G})\hat x^\dagger_{\mathbf{k}+\mathbf{G}}\hat x_\mathbf{k}
 \nonumber\\
+\frac1A\sum_{\substack{\mathbf{q}\in \text{BZ}\\ \lambda}}\sum_{{\mathbf G}\in \text{RL}} 
g_{\mathbf{q},\mathbf{G},\lambda}  ( \hat{b}_{\lambda,\mathbf{q}} + \hat{b}_{\lambda,-\mathbf{q}}^\dagger)
\sum_{\mathbf k}\hat{x}^\dagger_{\mathbf{k}+\mathbf{q}+\mathbf{G}} \hat{x}_{\mathbf{k}}.
\label{Hamiltonian}
\end{gather}
Here, $\hat x^\dagger_\mathbf{k}$ creates an exciton in the top plane with momentum $\mathbf{k}$ and energy $\epsilon_{\mathbf k}=k^2/2m_x$ with $m_x$  the exciton mass, and
$\hat{b}_{\lambda,\mathbf{q}}^\dagger$ creates a phonon 
in the WC in mode $\lambda=1,2$  with energy $\omega_{\mathbf{q}\lambda}$ and momentum $\mathbf{q}$ inside the first Brillouin zone (BZ) of the WC. 
$A$ is the area of the planes, and we have defined 
$V_\text{B}({\mathbf G})=A^{-1} \int\!d^2re^{-i{\mathbf G}\cdot{\mathbf r}}V_\text{B}({\mathbf r})$, where 
 $V_\text{B}({\mathbf r})=\sum_i V_\text{ex}({\mathbf r}-\mathbf{R}_i^0)$ is the periodic Bloch potential on the exciton from  the electrons in the equilibrium positions of the 
 WC and 
 % with equilibrium positions  $\mathbf{R}_i^0$ and reciprocal lattice (RL) vectors ${\mathbf G}$. 
 the integral $\int_\text{cell}\!d^2r$ is over its  primitive unit cell. 
 The exciton-phonon interaction vertex is 
\begin{equation}
  g_{\mathbf{q},\mathbf{G},\lambda}=-i\sqrt{\frac{N}{2m_e\omega_{\mathbf{q}\lambda}}}({\mathbf q}+{\mathbf G})\cdot\boldsymbol{\epsilon}_{\mathbf{q}\lambda}
  V_\text{ex}({\mathbf q}+{\mathbf G})
  \label{gVertex}
\end{equation}
where $m_e$ is the electron mass, $\boldsymbol{\epsilon}_{\mathbf{q}\lambda}$ is the phonon polarization vector,  $N$ is the number of sites (electrons) in the WC, 
 and   $V_\text{ex}({\mathbf q})= A^{-1}\int\!d^2re^{-i{\mathbf q}\cdot{\mathbf r}}V_\text{ex}({\mathbf r})$. This vertex gives from Eq.~\eqref{Hamiltonian} the strength 
 of the exciton absorbing/emitting a phonon with momentum $\pm{\mathbf q}$ in mode $\lambda$ with momentum conservation modulo 
 a reciprocal lattice vector.  We use units where $\hbar =1$.

\subsection{Bloch bands and phonon spectrum}
\begin{figure}
    \centering
    \includegraphics[width=\columnwidth]{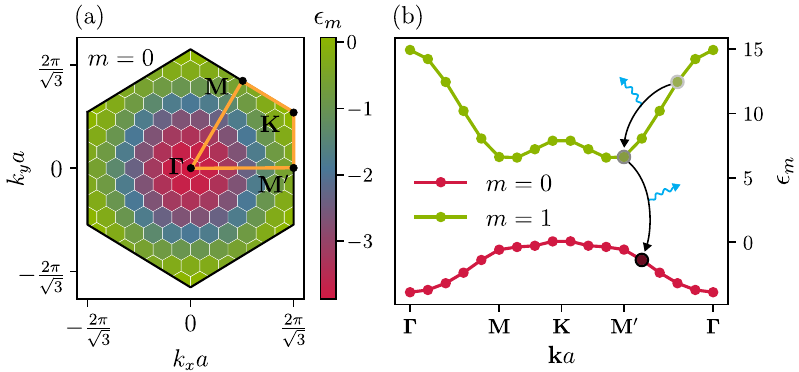}
    \caption{ (a)
    Contour plot of the lowest ($m=0$) Bloch band   for $d/a=0.05$,  $m_{x}a^2\kappa/d^4=217.5$, and a system size of $11\times11$ with energies  in units of  $1/m_xa^2$. (b)
    The two lowest Bloch bands following the orange path in the BZ shown in panel (a). The gray/black filled circles indicate inter- and intraband scattering via the emission of phonons (blue wavy lines).}
    \label{fig.DispTLB}
\end{figure}

The periodic potential $V_B$ in Eq.~\eqref{Hamiltonian} gives rise to  Bloch bands for the exciton with energies $\epsilon_{m\mathbf{q}}$ where $m$ is the band index and $\mathbf{q}\in \text{BZ}$. Writing the Bloch wave functions as  $\phi_{{\mathbf q}}^m({\mathbf r})=A^{-1/2}\sum_{\mathbf G}\phi_{{\mathbf q},{\mathbf G}}^me^{i({\mathbf q}+{\mathbf G})\cdot{\mathbf r}}$ we get 
\begin{equation}
\frac{(\mathbf{q}+\mathbf{G})^2}{2m_x} \phi^{m}_{\mathbf{q},\mathbf{G}} + \sum_{\mathbf{G'}}V_B(\mathbf{G}-\mathbf{G}')\phi^m_{\mathbf{q},\mathbf{G'}}
= \epsilon_{m\mathbf{q}}\phi^m_{\mathbf{q},\mathbf{G}}. 
\label{BlochEquation}
\end{equation}
Figure \ref{fig.DispTLB} shows the two lowest Bloch bands obtained numerically for  $d/a=0.05$, and $m_{x}a^2\kappa/d^4=217.5$.

The annihilation operator for an exciton with momentum ${\mathbf q+\mathbf G}$ where ${\mathbf q}\in \text{BZ}$ can be 
expressed in terms of annihilation operators of excitons in the Bloch bands as 
\begin{equation}
\hat x_{\mathbf q+\mathbf G}=\sum_m\phi_{{\mathbf q},{\mathbf G}}^m\hat x_{m\mathbf q}. 
\label{BlochCreation}
\end{equation}
Using Eq.~\eqref{BlochCreation}  in 
Eq.~\eqref{Hamiltonian} then gives a Hamiltonian describing Bloch excitons with energies  $\epsilon_{m\mathbf{q}}$, which can emit or absorb 
WC phonons with total crystal momentum conserved in two different ways: intraband scattering where the exciton stays in the same band or 
interband scattering where it changes its band.  For details, see Appendix~\ref{App.Ham}.
%We numerically solve for the Bloch eigenstate, thus diagonalizing $\hat{H}_X$ and $\hat{V}_{Bloch}$. $\hat{X}^\dagger_{m \mathbf{k}}$ create an exciton in the $m^{\text{th}}$ Bloch band with crystal momentum $\mathbf{k}$.  In Fig. \ref{fig.DispTLB}, we plot the two lowest Bloch bands for $\bar{\alpha}=30$meV and $a =23.5nm$. 
%%%%%%%%%%%%%%%%%%%%%%%%%%%%%%%%%%%%%%%%%%%%%%%%%%%%%%%%%%%%%%%%%%
%%%%%%%%%%%%%%%%%%%%%%%%%%%%%%%%%%%%%%%%%%%%%%%%%%%%%%%%%%%%%%%%%%

\begin{figure}
\centering
  \includegraphics[width=1\columnwidth]{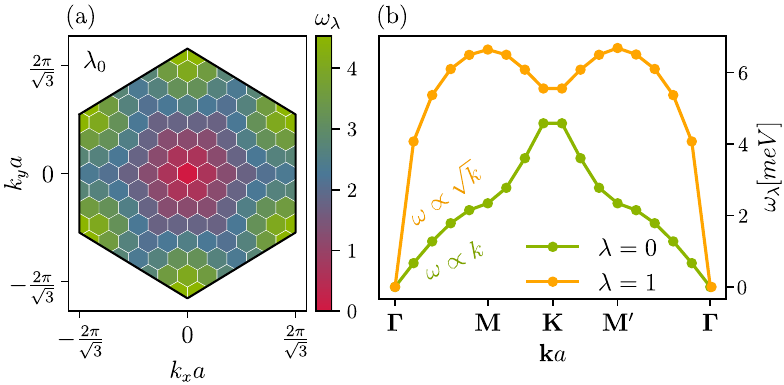}
\caption{(a) Contour plot of the lowest (transverse) phonon mode 
 for $a=23.5$nm and calculated using a system size of $11\times11$. (b) Dispersion of the transverse and longitudinal phonon modes along the orange path shown in Fig.\ \ref{fig.DispTLB}(a) with their 
  dispersion  around the $\mathbf{\Gamma}$-point indicated.}
\label{phononmodes}
\end{figure}

 The phonon modes are found from the usual harmonic approximation where we use Ewald resummation to numerically accelerate convergence of the resulting sums of 
 the long-range Coulomb interaction~\cite{Ashcroft76, Nelson2014}. We obtain two phonon branches, which recover the 
 long-wavelength dispersions  
 \begin{equation}
\omega_{ql}=\frac{2\sqrt\pi}{3^{1/4}} \omega_C\sqrt{qa}\hspace{0.5cm}\text{ and }\hspace{0.5cm}\omega_{qt}=\frac{2^{1/4}\eta^{1/2}}{3^{1/8}} \omega_Cqa
\label{eq.phononmodes}
 \end{equation}
for the transverse $(\lambda = 0)$ and the longitudinal $(\lambda = 1)$ modes with $\eta=0.25$. Here, 
%\jens{\sout{Here} Capturing the low-energy behavior and describing the scaling of the bandwidth}
\begin{equation}
    \omega_C=\sqrt{\frac{Q^2}{m_e a^3}}%\simeq 1.5 \text{meV}
    \label{eq.PhonE}
\end{equation} 
is  the characteristic phonon energy of the WC with $Q^2=e^2/4\pi \epsilon_r \epsilon_0$ ~\cite{Fisher1979}. It gives the frequency scale of a single electron oscillating in the WC while all the others are static in the 
spirit of the Einstein approximation of phonons, 
and it therefore sets the bandwidth of the phonon continuum corresponding to the wave solutions where all electrons in the crystal oscillate together. 
The $\sqrt q$ scaling of the longitudinal phonon mode comes from the long-range nature of the Coulomb interaction, making 
it energetically costly to excite. This suppresses the effects of the longitudinal mode on the exciton as compared to the transverse mode. 
In Fig.~\ref{phononmodes}, we plot the spectrum of the two phonon branches obtained numerically using an  $11\times11$ lattice. 
% \begin{figure}
% \centering
%   \includegraphics[width=0.89\columnwidth]{Contour_Bloch.pdf}
% \caption{Dispersion for the two types of phonon modes with $\lambda_{0} \propto p $ and $\lambda_{0} \propto \sqrt{p} $ around $p=0$. These are calculated for $a=17$nm, a system size of 11x11, and units of $\omega_{\lambda}$ is MeV. \georg{Use $\omega_C$ as energy unit. Then everything becomes parameter independent.  
% Plot longs retninger i BZ med kontur-plots som inds\ae t. }}
% \label{phononmodes}
% \end{figure}
A detailed discussion of the phonon mode calculation is given in Appendix~\ref{App.Ham}.

\subsection{Exciton-phonon interaction strength}
It is important to understand what governs the   strength of the interaction between the exciton and the Wigner phonons.
To obtain an expression for this,  we calculate 
the second-order energy shift of the exciton due to the emission and subsequent absorption of a phonon (note that there is no first-order energy shift since only even-order terms preserve the number of phonons).  
Including only the lowest linear phonon mode with velocity $\omega_Ca$ in Eq.~\eqref{eq.phononmodes}, this yields 
\begin{equation}
\Sigma_2\sim\int\! d^2q\frac{|g_{\bq}|^2}{A}\frac{1}{\omega_Caq}\sim %\frac{1}{m_ea^2}\frac{V_\text{ex}^2(1/a)}{\omega_C^2a^2}\sim 
\frac{1}{m_xa^2}\frac{m_x}{m_e\omega_C^2}
\frac{\kappa^2}{d^8}\frac{d^4}{a^4}=\frac{\gamma^2}{m_xa^2}.
\label{intstrength}
\end{equation}
Here we have used $N/A\sim 1/a^2$, that the Fourier transform scales as $V_\text{ex}(1/a)\sim -\kappa/d^2$, and that 
the typical kinetic energy of the exciton is much smaller than the typical phonon energy inside the BZ. 
Equation \eqref{intstrength} shows that the exciton-phonon interaction strength $\gamma^2$
relative to the bare kinetic energy of the exciton scales with the typical exciton-electron interaction energy over  a typical phonon energy squared. 
Since $1/\omega_C^2\propto a^3$, we have $\gamma^2\propto 1/a\propto \sqrt n$ explicitly showing how the exciton-phonon interaction increases with density. We shall see, however, that 
a counteracting effect is that the phonon spectrum becomes off-resonant with the exciton spectrum
for large densities thereby decreasing its effects.
%\jens{From this, we get that the interaction strength increases with density, but we also see that due to the behavior of the bandwidth as a function of density, the importance of the interactions actually decreases again a higher densities.}

\section{Quantum field theory} \label{sec.QFT}
To explore how the phonons modify the dynamics of the exciton, we compute the retarded exciton Green's function.
This is most naturally calculated in the Bloch basis of the excitons where it is defined as 
\begin{align}
	G_{mm'}(\bq,t) = -i\theta(t)\langle [ \hat{x}_{m\bq}(t), \hat{x}^{\dagger}_{m'\bq}(0)]\rangle,
	\label{eq.Green}
\end{align}
with $\hat O(t)=e^{iHt}\hat Oe^{-iHt}$  an operator in the Heisenberg picture. Note that the exciton Green's function is not diagonal in  the  Bloch band index 
 $m$ since  scattering on phonons can change the band of the excitons as discussed above. We will later rotate to the plane-wave
 basis probed by experiments. 
 
 We use a so-called  self-consistent Born approximation (SCBA) to calculate the exciton self-energy, which amounts to summing the 
 noncrossing "rainbow" Feynman diagrams to effectively infinite order. This approximation  is known to give a  remarkably 
 accurate description of holes in antiferromagnetic lattices~\cite{Kane1989,Martinez1991,Liu1991,Schmitt-Rink1988,KKNielsen2021} when  
  compared to Monte Carlo calculations~\cite{Diamantis_2021}. This accuracy was 
  recently shown to hold  even for nonequilibrium cases when compared  
 with optical lattice experiments~\cite{KKNielsen2022}, and the SCBA has also been generalized to   dopants in spin liquids~\cite{Nyhegn2025,nyhegn2025spinchargeboundstatesemerging}.  
 
\begin{figure}
    \centering
    \includegraphics[width=0.95\columnwidth]{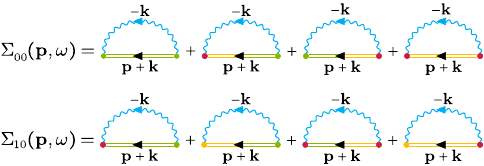}
    \caption{Feynman diagrams for the exciton self-energies associated with intraband $\Sigma_{00}$ and interlayer interband $\Sigma_{10}$ scattering. 
    Wavy lines indicate phonons  while the double lines are the dressed exciton propagator with the green/yellow color representing the $m=0$/$m=1$ Bloch bands.  The red vertex is associated 
    with scatterings where the exciton changes band, while the green and yellow vertices represent scatterings where the exciton stays in the same band. }
    \label{fig.Sigma}
\end{figure}

 We focus here on the effects of phonons on excitons in the two lowest   Bloch bands with $m=0$ and $1$ that  are most easily accessible experimentally. 
 Neglecting  the effects of higher bands to which to the coupling is weak then yields a  Hamiltonian with precisely the same structure as that 
 describing a  dopant in a bilayer  antiferromagnet, with the two Bloch bands corresponding to the two layers~\cite{Nyhegn2022,Nyhegn2023}. 
 The Green's function acquires a $2\times2$ matrix structure with $[\mathbf{G}(\bp,\omega)]_{mm'}= G_{mm'}(\bp,\omega)$, 
 and it obeys the Dyson 
 equation $\mathbf{G}(\bp,\omega) = \mathbf{G}_{0}(\bp,\omega) + \mathbf{G}_{0}(\bp,\omega) \boldsymbol{\Sigma}(\bp,\omega)  \mathbf{G}(\bp,\omega)$. Within the SCBA, the self-energy is  
\begin{align}
	\boldsymbol{\Sigma}(\bp,\omega) =  \sum_{\substack{\bk\in\text{BZ}\\\lambda}} {\bf g}^{\dagger}_{\lambda}(\bp,\bk) \boldsymbol{{\rm G}}(\bp + \bk,\omega - \omega_{\lambda,-\bk}) {\bf g}_{\lambda}(\bp,\bk)
\label{Eq:Self_Energies}
\end{align}
where the vertex matrix is defined by
\begin{align}
	\bf{g}_{\lambda}(\bp,\bk) = \begin{bmatrix} g^{00}_{\bp,\bk,\lambda}  & g^{01}_{\bp,\bk,\lambda}  \\ g^{10}_{\bp,\bk,\lambda}  & g^{11}_{\bp,\bk,\lambda}  \end{bmatrix}.
	\label{Eq:Coupling_Matrix} 
\end{align}
Here, $g^{mm'}_{\bp,\bk,\lambda}$ is the vertex describing an exciton with momentum $\bp$ in Bloch band $m$ scattering to momentum $\bp+\bk$ in band $m'$, by absorbing or emitting a phonon with momentum $\pm{\bp}$ in mode $\lambda$. We have used $g^{mm'}_{\bp+\bk,-\bk,\lambda} = (g^{m'm}_{\bp,\bk,\lambda})^{*}$, which also leads to $\Sigma_{01}=\Sigma_{10}$ and $G_{01}=G_{10}$. 
Two Feynman diagrams included in the calculating the self-energies in Eq.~\eqref{Eq:Self_Energies} are  shown in  Fig.\ \ref{fig.Sigma}.
The second order expression Eq.~\eqref{intstrength} corresponds to the Feynman diagram using a bare exciton propagator and ignoring Bloch band effects. 
The  self-consistent equation  for the Green's function is numerically solved iteratively starting from $\boldsymbol{\Sigma}=\mathbf{0}$ until convergence is reached. More details regarding our Green's function approach are given in  Appendix.~\ref{App.FullGreens}.

%%%%%%%%%%%%%%%%%%%%%%%%%%%%%%%%%%%%%%%%%%%%%%%%%%%%%%%%%%%%%%%%%%%%%%%%%%%%

%%%%%%%%%%%%%%%%%%%%%%%%%%%%%%%%%%%%%%%%%%%%%%%%%%%%%%%%%%%%%%%%%%%%%%%%%%%%

%Inserting the expressions for the Green's functions into Eq. \eqref{Eq:Self_Energies}, leads to a set of self-consistent equations we can solve 
%From this, we obtain expression for the full Green's function as a function of the self-energies, for expressi

\section{Results} \label{Sec.Res}
We now discuss our numerical results for the exciton spectral function. For clarity,  the spectral function is first analyzed in the exciton Bloch basis 
and then in the plane-wave basis that  is measured experimentally using 
optical spectroscopy~\cite{massignan2026}.

In the experimental setup, we envisage  using TMDs 
 such as MoS$_{2}$, MoSe$_{2}$, MoTe$_{2}$, WS$_{2}$ and WSe$_{2}$. This corresponds to 
 a relative permittivity $\epsilon_r$ between $4.35$ and $4.5$, an exciton radius $a_{x}$ between $1.1$ and $1.7$nm, and an exciton 
  polarizability $\alpha$ between $0.9 \cdot 10^{-36}$ and $1.45 \cdot 10^{-36}$ Cm$^{2}$/V \cite{goryca2019, zipfel2018, fey2020}. 
  We choose $d=a_{x}$ as the thickness of the isolating hBN layer so that the electron-exciton interaction is accurately described by 
  the dipole-charge interaction, Eq.~\eqref{potentialfull}.

%  This gives an estimate between $30$meV and $300$meV. To illuminate the role of phonons, we will vary $\bar{\alpha}$ between these estimates and keep  strength of the interaction to illuminate the role of phonons, but

\subsection{Bloch states  spectral function}\label{BlochSection}
\begin{figure}
    \centering
    \includegraphics[width=1\columnwidth]{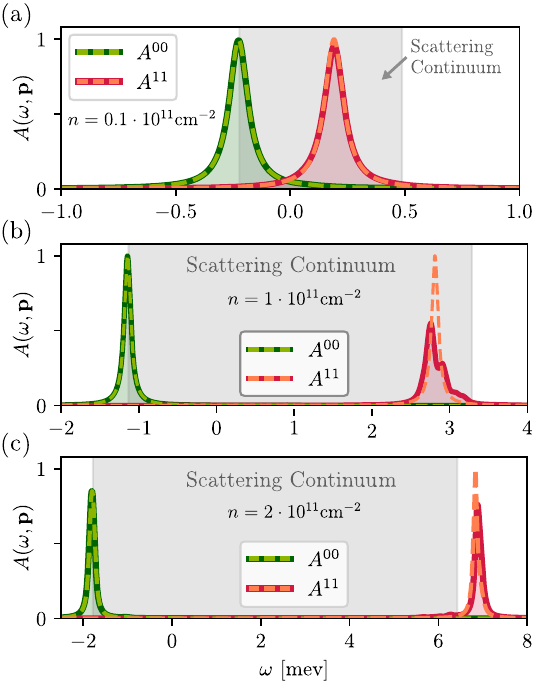}
    \caption{Diagonal exciton spectral functions in the Bloch band basis calculated with (full) and without (dashed) interband scattering for $\bp = \bf \Gamma $ and the indicated densities. The spectral functions are normalized such that the maximum value is $1$.
    }
    \label{fig.spec0}
\end{figure}

Figure \ref{fig.spec0} shows the diagonal exciton spectral functions $A_{mm}(\bp,\omega)=-2\text{Im}G_{mm}(\bp,\omega)$
for zero momentum $\bp=0$ and the lowest and first excited Bloch band $m=0,1$.
We have taken $\kappa/d^4=100$meV and three different electron densities of the WC: 
$n=0.1\times10^{11}$cm$^{-2}$, $n=1\times10^{11}$cm$^{-2}$, and $n=2\times10^{11}$cm$^{-2}$ corresponding 
to the exciton-phonon coupling strengths $\gamma=0.04$, $0.08$, and $0.1$ respectively. 
%the electron density of the WC $n=2\cdot10^{11} cm^{-2}$  corresponding to $d/a=\georg{?}$,  
% $(Q^2/a)/\omega_C=\georg{?}$, and the exciton-phonon interaction strength $(\kappa/d^4)/\omega_C=\georg{?}$. \jens{(Do we not confuse the reader, by listing all these quantities?)}
The   calculated spectral functions all frequency integrate to $2\pi$ within $0.3\%$ demonstrating the accuracy of our 
numerics. 

We see from Fig.~\ref{fig.spec0} that the ground-state $m=0$ exciton remains well defined, and we find that its energy  
 is slightly lower than the bare Bloch band energy  $\epsilon_{0,\mathbf{0}}$ obtained by 
solving Eq.~\eqref{BlochEquation} for all three densities. 
This is expected since this band has a minimum at the momentum $\bq=0$ (see Fig.~\ref{fig.DispTLB}), so that exciton can only virtually emit and absorb phonons giving rise to a redshift. 
 
Consider next the exciton in the excited $m=1$  Bloch band, which contrary to the ground-state exciton can decay to the lowest $m=0$ band by emitting a phonon.  This interband scattering event is resonant when the exciton has an energy between 
the minimum and maximum of $\epsilon_{0,-\bq} + \omega_{\lambda,\bq}$ indicated by the gray regions in Fig.~\ref{fig.spec0}. Also, since the $\boldsymbol{\Gamma}$-point is a local maximum for the $m=1$ Bloch band, the exciton can emit phonons while staying 
 in the $m=1$ band in an intraband scattering event.
 Both processes can give rise  to broadening and renormalization but we shall see that resonant interband scattering is the dominant channel. 

For the  case of low electron density $n=0.1\times10^{11}$cm$^{-2}$ shown in Fig.\ \ref{fig.spec0}(a), we see little broadening 
of the $m=1$ peak even though it is inside the interband  exciton-phonon scattering continuum. This is because   
the interaction strength $\gamma=0.04$ is small. %\georg{Beh\o ver vi vise panel (a)? Der sker jo ikke noget. }
%Inside this continuum, the exciton can resonantly emit phonons 
%with momenta $\bq$  so that $\epsilon_{1\mathbf{0}}=\epsilon_{0-\bq}+\omega_{\lambda\bq}$, but this does not give rise to significant damping because the interaction strength is so weak at low densities.
As the density increases, the magnitude of the exciton-phonon coupling grows, which in general leads to stronger 
renormalization and damping of the excitons. 
For example, for the density $n = 1\times 10^{11}\mathrm{cm^{-2}}$ shown in Fig.\ \ref{fig.spec0}(b)
corresponding to the coupling strength $\gamma=0.08$, the effects of phonons are significant and the $m=1$ exciton
is substantially broadened. To disentangle the effects of intra- and interband scattering, we plot as dashed lines the 
spectral function obtained when the interband coupling is omitted. This demonstrates that the broadening 
of the $m=1$ exciton is mainly due to resonant interband scattering, consistent with the fact that it is inside the 
interband scattering continuum. Figure \ref{fig.spec0}(c) shows that for 
 the even larger density $n = 2\times 10^{11}\mathrm{cm^{-2}}$ corresponding to $\gamma=0.1$, the 
$m=1$ exciton  is, however, again well-defined. 
The reason is that its energy  is outside the interband-scattering continuum so that there is no resonant scattering. 
This trend arises because the characteristic phonon energy scales as $\omega_{C} \propto a^{-3/2}\propto n^{3/4}$, whereas the energy of
the excited $m=1$ Bloch exciton scales more rapidly as $\epsilon_{1,\mathbf{q}=0} \propto a^{-2}\propto n$. Consequently, even though the 
magnitude of the exciton-phonon coupling strength $\gamma^2$  increases with density, the energy separation between the $m=1$ and $0$
exciton Bloch states increases faster. Eventually, the $m=1$ exciton moves outside the resonant interband scattering continuum making it 
well-defined again. We note that when the density becomes too high, the WC eventually melts into a Fermi liquid again.

\subsection{Plane-wave spectral function}
Optical  spectroscopy measures  the exciton spectral function in the plane-wave basis for ${\bf k}=0$ rather than in the Bloch basis. Using 
 Eq.\ \eqref{BlochCreation}, the exciton plane-wave spectral function can be expressed in terms of its Green's function in the Bloch basis as 
\begin{align}
%	G^{P}(\bk + \bG,\omega) &= \sum_{mm'} [\phi^{m\mathbf{k}}_{\mathbf{G}}]^{*}\phi^{m'\mathbf{k}}_{\mathbf{G}} G_{mm'}(\bk,\omega) \nonumber \\
	%A^{P}(\bk + \bG,\omega) &= -2\sum_{mm'} \text{Im}\left[(\phi^{m\mathbf{k}}_{\mathbf{G}})^{*}\phi^{m'\mathbf{k}}_{\mathbf{G}} G_{mm'}(\bk,\omega)\right],\nonumber \\
    A^{P}(\bk + \bG,\omega) &= -2\text{Im}\sum_{mm'} (\phi^{m\mathbf{k}}_{\mathbf{G}})^{*}\phi^{m'\mathbf{k}}_{\mathbf{G}} G_{mm'}(\bk,\omega)
	\label{eq.PWGreen}
\end{align}
where $\bk\in$ BZ. This shows that the umklapp peaks at the experimentally relevant $\Gamma$ point $\bk + \bG = \mathbf{0}$ are proportional to $|\phi^{m\mathbf{0}}_{\mathbf{0}}|^2$, which will decrease with 
increasing electron density as the energy spacing between the Bloch bands increases in agreement with what is observed experimentally~\cite{Smolenski2021}. 

\begin{figure}
    \centering
    \includegraphics[width=1\columnwidth]{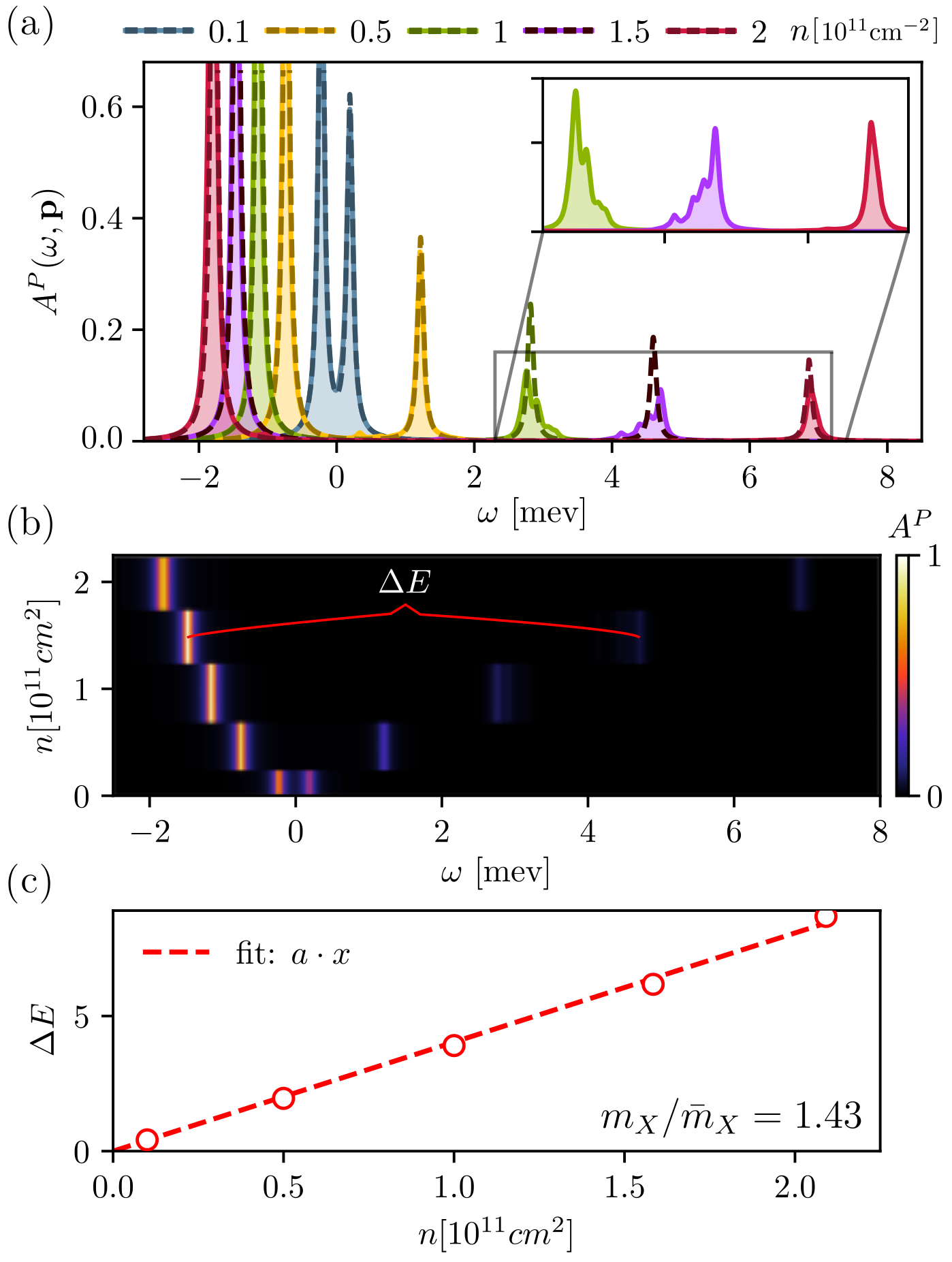}
    \caption{
 (a) Exciton spectral functions in the plane-wave basis for $\bp = \bf \Gamma$ and different densities. The full (dashed) lines are the results with (without) phonon scattering, 
 and the inset shows a zoomed-in section without the dashed lines. (b) Contour plot of the  $\bp = \bf \Gamma$  plane-wave spectral as a  function of the electron density and energy. (c) The energy 
 difference $\Delta E$ between the lowest and the first excited state as a function of electron density together with a  linear fit yielding an exciton mass $m_x=1.43\bar{m}_{x}$ where $\bar{m}_{x}$ is the bare exciton mass. The spectral functions are normalized such that maximum intensity is $1$.}
    \label{fig.specPW}
\end{figure}

When phonons are included, the behavior of the umklapp peaks becomes more intricate. In Fig.~\ref{fig.specPW}(a), we plot the exciton
plane-wave spectral function at the $\mathbf{\Gamma}$-point with (solid lines) and without (dashed lines) phonons for several densities. This comparison highlights the nontrivial effects of the phonons. 
At low densities $n \lesssim 0.5 \times 10^{11}\mathrm{cm}^{-2}$, phonons have only small effects and both the ground state and the umklapp peak are well described  by the mean-field Bloch theory. 
At intermediate densities  $0.5 \times 10^{11}\mathrm{cm}^{-2} \lesssim n \lesssim 2 \times 10^{11}\mathrm{cm}^{-2}$, on the other hand, phonons have significant effects and the umklapp peak 
is substantially broadened. At higher densities still, there is again only small broadening since the interband scattering is no longer resonant, consistent with what we found in Sec.~\ref{BlochSection}. 

To more clearly illustrate the spectral weight of the ground state and umklapp peaks, we present in Fig.~\ref{fig.specPW}(b) a density plot of the plane-wave exciton spectral function at the 
$\mathbf{\Gamma}$-point. This highlights how the umklapp peak is strongly suppressed with increasing electron density and also reveals the energy shifts induced by the  dipole-electron interaction Eq.~\eqref{potentialfull}. 
By numerically locating the poles of the spectral function, we extract the energy difference $\Delta E$ between the ground state and first excited umklapp peak  as shown in Fig.~\ref{fig.specPW}(c). 
Fitting this to the formula $\Delta E =|\mathbf{G}|^{2}/2m_{x} = n/\sqrt{3}m_{x}$ yields 
that exciton-electron interactions and phonon scattering  primarily lead to a renormalization of the exciton effective mass $m_x$ compared to its bare mass $\bar m_x$ so that $m_x/\bar m_x\simeq1.43$,
while preserving the linear dependence predicted by the Bloch picture.

We note that all these results of course depend on the strength of the exciton-electron interaction Eq.~\eqref{potentialfull}, and we expect that a larger polarizability will lead to even stronger effects of the phonons.

%%%%%%%%%%%%%%%%%%%%%%%%%%%%%%%%%%%%%%%%%%%%%%%%%%%%%%%%%%%%%%%%
%%%%%%%%%%%%%%%%%%%%%%%%%%%%%%%%%%%%%%%%%%%%%%%%%%%%%%%%%%%%%%%%

% \begin{align}
%     \sum_{m}[\phi^{m\mathbf{k}}_{\mathbf{G}}]^{*} \hat{X}_{m \mathbf{k}} = \hat{X}_{ \mathbf{k+G}}
%     \label{eq.EigVec_rev}
% \end{align}

\section{Conclusions and outlook}
 We explored the properties of an exciton interacting with an electronic Wigner crystal 
including the coupling to its gapless phonon mode spectrum. The significance of the exciton-phonon coupling was shown to be determined 
by the  exciton-electron interaction strength relative  to the typical phonon energy squared. Using a field-theoretic approach 
based on a nonperturbative 
self-consistent Born approximation, we demonstrated that the scattering processes where the exciton emits or absorbs a 
phonon while staying within the same Bloch band or changing to a new one leads to the formation of polaronic quasiparticles.
These polarons  consist of the exciton in a Bloch state dressed by phonons, and their properties depend on the electron density in a 
nontrivial way. In particular, since the exciton and phonon dispersions depend on the electron density in different ways, we 
showed that the exciton is inside the gapless phonon scattering continuum  giving rise to strong damping only at intermediate electron densities. 
Finally, we discussed how these results influence  the   exciton spectrum observed experimentally. 

In our analysis, we assume that screening effects can be captured by an effective dielectric constant $\epsilon_r$.
This is because the screening exhibits only a weak frequency dependence at the relevant energy scales, as it originates from electrons in filled valence bands of the TMD layer and in the encapsulating hBN layers, both of which possess large band gaps. Including any such weak frequency-dependence will likely change the long-wavelength phonon dispersion $\omega \propto \sqrt{q}$ to a
 linear $\omega \propto q$ behavior for small $q$. This will in turn increase the phonon density of states 
 at low energies leading to a slightly stronger
 broadening of the umklapp peak. % due to an increased phase space for low-energy scattering.

It would be interesting to include more intricate exciton band structures in our analysis, which, for instance, can 
arise  due to the coupling of K and K' valley excitons in TMDs, as in the pioneering
experiment observing an electronic  Wigner crystal~\cite{Smolenski2021}. In  this experiment, 
%Another experimentally relevant question in this context concerns the role of phonons when
the exciton moreover resides in the same layer as the WC, which can lead to bound-state (trion) formation and thus additional quasiparticle branches beyond the 
repulsive polaron. Our results are in this sense primarily relevant for the repulsive polaron branch observed in these experiments, where the exciton is  dressed by electron-hole excitations with no bound state formation. It could be very interesting to include the possibility of bound state formation in 
our formalism so that we can explore how this is affected by phonon dressing.
When the exciton is in the same layer as the WC,  short-range corrections to the exciton-electron interaction to the charge-dipole form may become important
as the exciton can get close to the electrons. 
However, we expect such short-range corrections to lead only to small quantitative effects, since the lattice spacing of the Wigner crystal is much larger than the exciton size. 
%A unified description that includes both bound-state formation and phonon scattering remains an open problem.
In particular, 
the damping window will remain essentially unchanged since %we also expect this to persist when the exciton is in the same layer as the WC. This is because 
it arises from the interplay between the phonon bandwidth set by $\omega_{C}$ and the interband energy scale scaling as $1/a^{2}$, which are insensitive to the detailed 
form of the exciton-electron interaction.
%our conclusions regarding the existence of the damping window remain qualitatively unchanged for different interaction types.

Beyond the single-polaron properties, one could explore additional features of Bloch polarons emerging from exciton-WC coupling, such as their transport coefficients~\cite{erkensten2025impactelectronwignercrystal}, 
induced interactions~\cite{Paredes2024}, and  possible many-body states for nonzero densities. Finally, a fascinating research direction concerns how exciton spectroscopy can serve as a quantum probe for 
other strongly correlated  electronic states with or without density order predicted to exist in 
TMDs~\cite{massignan2026,Mak2022}.

{\it Note added.} In another independent work, the properties of an exciton in the same layer as the  WC leading to bound state (trion) formation were analyzed~\cite{adlong2025theoryexcitonpolarons2d}. Also, two 
new experimental results regarding polaron formation with excitons in a WC have appeared~\cite{zhang2025wignerpolaronsrevealwigner,wang2025spectroscopywignercrystalpolarons}.

{\it Acknowledgments.} We acknowledge E.\ Dizer and A.\ Christianen for  useful discussions and for pointing out Refs.~\cite{adlong2025theoryexcitonpolarons2d,zhang2025wignerpolaronsrevealwigner,wang2025spectroscopywignercrystalpolarons}. We also thank P.\ Massignan, 
T.\ Pohl and K.\ M\o lmer  for insightful discussions. Financial support is provided by the Villum Foundation, the Independent Research Fund Denmark-Natural Sciences via grant No. DFF -8021-00233B, US Army CCDC Atlantic Basic and Applied
Research via grant W911NF-19-1-0403, the Carlsberg Foundation (grant ID CF25-0422), and the Independent Research Fund Denmark (grant ID 10.46540/5341-00014B). 
\newpage
%\onecolumngrid
\appendix

\section{Derivation of the Hamilton} \label{App.Ham}

In this Appendix we present the origin of the Fr\"olich Hamilton, Eq. \eqref{Hamiltonian}. Writing the Hamilton as 
\begin{align}
    \hat{H}=\hat{H}_X + \hat{H}_{\text{Ph}} + \hat{V}_{\text{Int}}, 
\end{align}
we have that $\hat{H}_X$ and $\hat{H}_{\text{Ph}}$ describe the dispersion of the exciton and phonons respectively, while $\hat{V}_{\text{Int}}$ describes the interactions between the exciton and the electrons in the Wigner crystal. Assuming a quadratic dispersion of the exciton around the $\Gamma$-point, we approximate the exciton's dispersion as
\begin{gather}
    \hat H_{X} = \sum_{\mathbf{p}} \epsilon_{\mathbf k} \hat x^\dagger_\mathbf{p}\hat x_\mathbf{p},
\end{gather}
with $\epsilon_{\mathbf p}=p^2/2m_x$. This bare dispersion effectively describes an exciton living in free space with mass $m_{x}$. The validity of this approximation comes from the lattice constant of the exciton layer being $\sim 0.1$nm \cite{reshak2003}, while it is $a \sim 10$nm for the Wigner crystal. The large difference means that the FBZ coming from the umklapp scattering of the Wigner crystal will be centered around the $\Gamma$-point such that only the high-lying Bloch bands will deviate from the quadratic dispersion.

%the layer with the ex this  Since the exciton lives in a lattice, this approximation will deviate for crystal fare Derivations from the approximation become increasingly important as the density of the Wigner crystal increases. Increasing density means a decreasing lattice constant, which leads to umklapp scattering between larger and larger momenta. A larger reciprocal lattice means that states around the $\Gamma$-point can scatter to states where the quadratic dispersion approximation does not apply. The lattice spacing of the exciton layer is $\sim 1$\AA  \cite{reshak2003}, while the lattice spacing of the Wigner crystal is $a = \sqrt{2/(\sqrt{3}n)} \sim 10^{3}$\AA$ $ causing umklapp scattering around the $\Gamma$-point to occur in its vicinity. If the scattering is not too strong, a quadratic dispersion of the exciton's dispersion around the $\Gamma$-point, hence, means this approximation works well \jens{reformulate}.

To retrieve the dispersion of the phonons, we apply the usual harmonic approximation where one solves a matrix equation of the form \cite{Ashcroft76, Nelson2014}
\begin{align}
    m_e^*\omega^2 \bm{\epsilon}(\mathbf{k})=\mathbf{D}(\mathbf{k})\bm{\epsilon}(\mathbf{k}),
    \label{eq.Phonon}
\end{align}
where $\mathbf{D}(\mathbf{k})=\sum_{\mathbf{R}} \mathbf{D}(\mathbf{R}) \exp(-\mathbf{k}\cdot \mathbf{R})$ is the Fourier transform of $\mathbf{D}(\mathbf{R})$ which in turn is given as 
\begin{align}
    \mathbf{D}_{ij}(\mathbf{R}) = \frac{e^2}{4\pi \epsilon_0 \epsilon_r} \times \begin{cases}
        \sum_{\mathbf{R}\neq \mathbf{0}} \left[ \frac{3R_i R_j }{R^5} - \frac{\delta_{ij} }{ R^3}\right], & \ \mathbf{R=0}, \vspace{0.1cm}\\
        \frac{\delta_{ij} }{R^3} - \frac{3 R_i R_j}{R^5}, & \ \mathbf{R} \neq \mathbf{0},
    \end{cases}
\end{align}
where we have assumed that the electrons in the Wigner lattice interact with each other via the repulsive Coulomb potential $V(r) = \frac{e^2}{4\pi \epsilon_0 \epsilon_r r} $. We can further write $\mathbf{D}(\mathbf{k})$ as  
\begin{equation}
    \mathbf{D}_{ij}(\mathbf{k}) = \frac{e^2}{4\pi \epsilon_0 \epsilon_r} \lim_{u\rightarrow 0} \frac{\partial^2}{\partial u_i \partial u_j} \sum_{\substack{n\\ \mathbf{R}_n\neq \mathbf{0}} } \frac{1}{|\mathbf{R}_n+\mathbf{u}_n|}\left(1-e^{-i\mathbf{k}\cdot\mathbf{R}_n}\right),
\end{equation}
where $\mathbf{u}_n$ is the displacement vector from equilibrium at lattice site $n$ as the static lattice approximation breaks down, so that the actual positions of the electrons in the Wigner lattice can be modeled as $\mathbf{R}_n=\mathbf{R}^0_n+\mathbf{u}_n$. From here, one can split the integrals in to short-range and long-range parts so that one can write $\mathbf{D}(\mathbf{k}) = \mathbf{D}^{>}(\mathbf{k})+\mathbf{D}^{<} (\mathbf{k})$, and these fast converging sums are readily computed numerically to obtain the phonon dispersion. With $\mathbf{D}(\bk)$, we solve Eq. \eqref{eq.Phonon} to find the dispersion of the two phonons, $\lambda=1,\ 2$, which in the long-wavelength limit go as $\omega_{\mathbf{q},1} \propto q$ and $\omega_{\mathbf{q},2}\propto \sqrt{q}$.

For the interacting part of the Hamilton, $\hat{V}_{\text{Int}}$, we start by expanding the electron-exciton potential to get  
\begin{align}
    V_{eX}(\mathbf{r}-\mathbf{R}_i) \approx V_{eX}(\mathbf{r}-\mathbf{R}^0_i)-\nabla_\mathbf{r} V_{eX}(\mathbf{r}-\mathbf{R}^0_i)\cdot \mathbf{u}_i.
\end{align}
This encourages the decomposition of the interaction as 
\begin{align}
    \hat{V}_{\text{Int}}=\hat{V}_{\text{Bloch}} + \hat{V}_{X-\text{Ph}},
        \label{intrealspace}
\end{align}
where 
\begin{align}
    \hat{V}_{\text{Bloch}} =  \int d^2 \mathbf{r} \hat{\Psi}^\dagger_X(\mathbf{r})\left(\sum_i V_{eX}(\mathbf{r}-\mathbf{R}^0_i) \right) \hat{\Psi}_X(\mathbf{r}),
\end{align}
\begin{align}
    \hat{V}_{X-\text{Ph}} =  -\int d^2 \mathbf{r} \hat{\Psi}^\dagger_X(\mathbf{r})\left(\sum_i \nabla_\mathbf{r} V_{eX}(\mathbf{r}-\mathbf{R}^0_i)\cdot \mathbf{u}_i \right) \hat{\Psi}_X(\mathbf{r}).
    \label{eq.VXPh}
\end{align}
To retrieve the Bloch equations, Eq. \eqref{BlochEquation}, we expand the electron-exciton potential in plane-waves
\begin{align}
    & \nonumber V_{eX}(\mathbf{r}-\mathbf{R}^0_j)  = \sum_{\mathbf{p}\in RS} V_{eX}(\mathbf{p}) e^{i\mathbf{p}\cdot (\mathbf{r}-\mathbf{R}^0_j)} \\ &= \sum_{\mathbf{k}\in \text{FBZ}}\sum_{\mathbf{G}\in \text{RL} } V_{eX}(\mathbf{q}+\mathbf{G}) e^{i(\mathbf{k}+\mathbf{G})\cdot (\mathbf{r}-\mathbf{R}^0_j)},
\end{align}
where we rewrote $\mathbf{p}=\mathbf{k}+\mathbf{G}$ for $\mathbf{q}\in \text{FBZ}$, and $\mathbf{G}$ is a reciprocal lattice vector. By inserting this into $\hat{V}_{\text{Bloch}}$, and expanding the real space exciton operators in terms of plane-waves, $\hat{\Psi}_X(\mathbf{r}) = \sum_{\mathbf{q}} \hat{X}_{\mathbf{q}} e^{i\mathbf{q}\cdot \mathbf{r}}$, we retrieve the Bloch.

To make progress on the exciton-phonon part, we note that
\begin{align}
    &\nabla_\mathbf{r} V_{eX}(\mathbf{r}-\mathbf{R}^0_j) = \sum_{\mathbf{q}\in \text{FBZ}} \sum_{\mathbf{G}\in \text{RL}} i (\mathbf{q}+\mathbf{G}) V_{\mathbf{q}+\mathbf{G}} e^{i (\mathbf{k+G})\cdot (\mathbf{r}-\mathbf{R}^0_j)},
\end{align} 
and quantizing the displacement in the standard way \cite{Bruus04}
\begin{align}
    \mathbf{u}_j =\frac{1}{\sqrt{N}} \sum_{\mathbf{k}\in \text{FBZ}}\sum_\lambda \frac{l_{\mathbf{k}\lambda} }{\sqrt{2}} \left( \hat{b}_{\lambda,\mathbf{q}} + \hat{b}_{\lambda,-\mathbf{q}}^\dagger \right) \mathbf{\epsilon}_{\mathbf{q} \lambda }e^{i\mathbf{q}\cdot \mathbf{R}^0_j},
\end{align}
with $N$ number of electrons in the WC, $l_{\mathbf{k}\lambda} = \sqrt{\hbar/(m_e^* \omega_{\mathbf{k}\lambda})}$, with $\omega_{\mathbf{k}\lambda}$ the phonon dispersion, and $\boldsymbol{\epsilon}_{\mathbf{q},\lambda}$ the phonon polarization vector. Using $\sum_j e^{i(\mathbf{k}-\mathbf{q})\cdot \mathbf{R}^0_j}=N\delta_{\mathbf{q},\mathbf{k}}$, we obtain
\begin{multline}
    \sum_i \nabla_\mathbf{r} V_{eX}(\mathbf{r}-\mathbf{R}^0_i)\cdot \mathbf{u}_i \\ = \sum_{\mathbf{q}\in \text{FBZ}, \lambda} \sum_{\mathbf{G}\in \text{RL}} g_{\mathbf{q},\mathbf{G},\lambda}\left( \hat{b}_{\mathbf{q},\lambda} + \hat{b}_{-\mathbf{q},\lambda}^\dagger \right)e^{i(\mathbf{q+G})\cdot \mathbf{r}},
\end{multline}
where
\begin{align}
    g_{\mathbf{q},\mathbf{G},\lambda} = i \sqrt{\frac{N\hbar}{2 m_e^* \omega_{\mathbf{q},\lambda} }}(\mathbf{q}+\mathbf{G})\cdot \boldsymbol{\epsilon}_{\mathbf{q},\lambda} V_{eX}(\mathbf{q}+\mathbf{G}),
    \label{gq}.
\end{align}
Again, expanding the exciton operators in terms of plane-wave operators in Eq. \eqref{eq.VXPh} we find a Hamiltonian of the form 
\begin{align}
    \frac{1}{A}\sum_{\substack{\mathbf{q}\in \text{BZ}\\ \lambda}}\sum_{{\mathbf G}\in \text{RL}} 
g_{\mathbf{q},\mathbf{G},\lambda}  ( \hat{b}_{\lambda,\mathbf{q}} + \hat{b}_{\lambda,-\mathbf{q}}^\dagger)
\sum_{\mathbf p}\hat{x}^\dagger_{\mathbf{p}+\mathbf{q}+\mathbf{G}} \hat{x}_{\mathbf{p}},
\label{eq.Vint}
\end{align}
where $\bp$ is a plane-wave momentum and can be decomposed as $\bp = \bk +\bG$ with $\bk \in \text{FBZ}$ and $\bG$ is a reciprocal-lattice vector. $A$ is the area of the Wigner crystal. Equation \eqref{eq.Vint} shows that the vertex $g_{\mathbf{q},\mathbf{G},\lambda}$ describes the scattering of a plane-wave exciton with momentum $\bp$ to $\bp+\bq+\bG$ by exiting(annihilating) a $\lambda$-phonon with crystal momentum $-\bq$($\bq$). 

Describing the excitons in terms of the Bloch states, $\hat x_{\mathbf k+\mathbf G}=\sum_m\phi_{{\mathbf k},{\mathbf G}}^m\hat x_{m\mathbf k}$, that diagonalize $\hat{H}_{X} + \hat{V}_{\text{Bloch}}$, we retrieve the final expression for the Hamilton:

\begin{align}
  \hat{H} &=  \sum_{m\mathbf{k}} \epsilon_{m,\mathbf k} \hat{x}^\dagger_{m,\mathbf{k}}\hat x_{m,\mathbf{k}} + \sum_{\lambda\mathbf{q}} \omega_{\lambda,\mathbf k} \hat{b}^\dagger_{\lambda,\mathbf{k}}\hat{b}_{\lambda,\mathbf{k}} \nonumber \\
  &- \frac{1}{A}\sum_{ \substack{m m' \\ \lambda} } \sum_{\mathbf{qk}}\tilde{g}^{m m'}_{\mathbf{k},\mathbf{q},\lambda} \hat{X}^\dagger_{m \mathbf{k}+\mathbf{q}} \hat{X}_{m' \mathbf{k}} \left( \hat{b}_{\lambda,\mathbf{q}} + \hat{b}_{\lambda,-\mathbf{q}}^\dagger \right),
\end{align}
with
\begin{align}
  &  \tilde{g}^{m m'}_{\mathbf{k'},\mathbf{q},\lambda}  \equiv \sum_\mathbf{G}g_{\mathbf{q},\mathbf{G},\lambda}\sum_{\mathbf{G}'} \phi^{m'}_{\mathbf{k},\mathbf{G'}}\left[\phi^{m}_{\mathbf{k} +\mathbf{q},\mathbf{G}+\mathbf{G'} }\right]^{*}
\end{align}
describing intraband ($m=m'$) and interband scattering $(m\neq m')$, and $\epsilon_{m,\mathbf k}$ is the dispersion of the $m$th Bloch band found by solving the Bloch equations Eq. \eqref{BlochEquation}

\section{Full expression for the Green's function} \label{App.FullGreens}
Obtained using the Dyson equation from Sec. \ref{sec.QFT} and Eq. \eqref{Eq:Self_Energies}, we state the full expression for the Green's functions:

\begin{widetext}
    \begin{align}
    G_{00}(\bp,\omega) &= \frac{  \omega - \epsilon_{1,\bp} - \Sigma_{11}(\bp,\omega) +i\eta}{ (\omega - \epsilon_{0,\bp} - \Sigma_{00}(\bp,\omega) +i\eta)(\omega - \epsilon_{1,\bp} - \Sigma_{11}(\bp,\omega) + i\eta) - \Sigma_{10}(\bp,\omega)\Sigma^{01}(\bp,\omega)} \nonumber \\
    G_{11}(\bp,\omega) &= \frac{  \omega - \epsilon_{0,\bp} - \Sigma_{00}(\bp,\omega) +i\eta}{ (\omega - \epsilon_{0,\bp} - \Sigma_{00}(\bp,\omega) +i\eta)(\omega - \epsilon_{1,\bp} - \Sigma_{11}(\bp,\omega) + i\eta) - \Sigma_{10}(\bp,\omega)\Sigma_{01}(\bp,\omega)} \nonumber \\
    G_{10}(\bp,\omega) &= \frac{ \Sigma_{10}(\bp,\omega) }{ (\omega - \epsilon_{0,\bp} - \Sigma^{00}(\bp,\omega) +i\eta)(\omega - \epsilon_{1,\bp} - \Sigma^{11}(\bp,\omega) + i\eta) - \Sigma_{10}(\bp,\omega)\Sigma^{01}(\bp,\omega)} \nonumber \\
    G_{01}(\bp,\omega) &= \frac{ \Sigma_{01}(\bp,\omega) }{ (\omega - \epsilon_{0,\bp} - \Sigma_{00}(\bp,\omega) +i\eta)(\omega - \epsilon_{1,\bp} - \Sigma_{11}(\bp,\omega) + i\eta) - \Sigma_{10}(\bp,\omega)\Sigma^{01}(\bp,\omega)}, 
    	\label{Eq:Prop}
    \end{align}
\end{widetext}

% \begin{figure}
%     \centering
%     \includegraphics[width=0.65\columnwidth]{Sigma2.pdf}
%     \caption{Feynman diagrams included when calculating the self-energy associated. (a) shows a diagram associated with interlayer propagation, while (b) shows a diagram associated with intralayer propagation. The curvy lines illustrates the phonons while the double full lines represent the dressed exciton propagator. The green exciton propagators represent propagation in Bloch band $m=0$ and the yellow the excited $m=1$ band. The red vertex is associated with interactions where the exciton changes band, while the green en yellow vertex represent interactions where the exciton continuous in the same layer. \georg{Switch back to old figure with all 
%     diagrams.}}
%     \label{fig.Sigma_old}
% \end{figure}

\bibliography{references}

%%%%%%%%%%%%%%%%%%%%%%%%%%%%%%%%%%%%%%%%%%%%%%%%%%%%%%%%%%%%%%%%%%%%%%%%%%%%%%%%%%%%%%%%%%%%%%

\clearpage

\end{document}